\documentstyle[prl,aps,epsf,psfig]{revtex}
\begin{document}
\twocolumn[\hsize\textwidth\columnwidth\hsize
           \csname @twocolumnfalse\endcsname
\title{Charge profile in vortices}
\author{Jan Kol\'a\v cek, Pavel Lipavsk\'y}
\address{Institute of Physics, ASCR, Cukrovarnick\'a 10,
16253 Prague 6, Czech Republic}
\author{Ernst Helmut Brandt}
\address{Max-Planck-Institut f\"ur Metallforschung,
         D-70506 Stuttgart, Germany}
\maketitle
\begin{abstract}
The electric charge density in the vortex lattice of superconductors
is studied within the Ginzburg-Landau theory. We show that the
electrostatic potential $\varphi$ is proportional to the GL function,
$\varphi\propto|\psi|^2-|\psi_\infty|^2$. Numerical results for the 
triangular vortex lattice are presented.
\end{abstract}
    \vskip2pc]
   Abrikosov vortices in type-II superconductor consist of
magnetic flux encircled by a rotating condensate of Cooper pairs.
The inertial and Lorentz forces both act in the centrifugal direction
and thus cause a small depletion
of the charges in the vortex core. The electrostatic potential
generated by this depleted charge supplies a centripetal force
which balances the centrifugal force \cite{London,LeBlanc}.

While the depleted charge occupies a region comparable with the
vortex core, i.e. a region extending over the Ginzburg-Landau (GL)
coherence length $\xi$, the screening charge extends over a region
extending over half the London penetration depth $\lambda$. This
charge distribution results in an electrostatic field which may
be observed either by nuclear quadrupole resonance \cite{Matsuda} or
by surface scanning as proposed recently by Blatter {\it et al.} 
\cite{Blatter} or by the Torricelli-Bernoulli effect \cite{Mishonov}. 
Besides, the charge of the vortex core is expected to contribute 
to the forces acting on the vortex. For example, Feigel'man 
{\it et al.} \cite{Feigel} speculate that this charge can explain 
the still puzzling sign reversal of the Hall voltage.

The charge in the asymptotic region away from the vortex core has been
studied by LeBlanc \cite{LeBlanc}. In this region one can benefit from
the simplicity of the London theory. Within the London theory, the
local balance of forces acting on the Cooper pair may be written in
form of the Newton equation,
  \begin{equation}   
  m^*\dot{\bf v}=-e^*\nabla\varphi+e^*{\bf v}\times{\bf B} \,.
  \label{Ne}
  \end{equation}
Here the mass $m^*=2m$ and charge $e^*=2e$ are twice the values of a
single electron. Unlike in common applications of the Newton equation
(\ref{Ne}) here the velocity is given by the London relation,
$m{\bf v}=-e{\bf A}$, and Eq.~(\ref{Ne}) thus specifies the electric
field $-\nabla\varphi$ which maintains this motion. 
Using the identity $\dot{\bf v} = ({\bf v}\nabla){\bf v} =
{1\over 2}\nabla\left({\bf v}^2\right)-{\bf v}\times\nabla
\times {\bf v}$ and the London 
approximation $e{\bf B}=-m\nabla\times {\bf v}$, 
one can show that Eq.~(\ref{Ne}) is solved
by the Bernoulli potential,
  \begin{equation}  
  \varphi=-{mv^2\over 2e} \,.
  \label{Bp}
  \end{equation}
The quadratic dependence on the velocity shows that the electric
potential $\varphi$ decreases over a characteristic length
$\lambda/2$. The charge density $\rho=-\varepsilon\nabla^2\varphi$
behaves similarly.

Obviously, the Bernoulli potential (\ref{Bp}) is not suited for the
core region, where the local London approximation fails. Moreover,
close to the core the velocity diverges,
$v\sim{e\Phi_0\over 2\pi m}{1\over r}$, where
$\Phi_0$ is the quantum of flux and $r$ is the distance from the
vortex axis. According to Eq.~(\ref{Bp}) and
$\rho=-\varepsilon\nabla^2\varphi$, the total screening charge
follows this divergence, $\int_r^\infty{\rm d}r \rho\propto r^{-2}$.
The charge of the core estimated from the London expression (\ref{Bp})
thus crucially depends on the estimate of the core diameter and the
profile of the charge is incorrect in the core region.

In the core region, a non-local approach like the Ginzburg-Landau theory
is required \cite{ehb1}. Like the London approach, the Ginzburg-Landau
theory does not include the electrostatic potential explicitly. Here we
show how this potential can be determined from its effect on the motion
of Cooper pairs. The Bernoulli potential then has to be replaced by the
potential
  \begin{equation}  
  \varphi={\beta\over 2e}\left(|\psi|^2-|\psi_\infty|^2\right) \,,
  \label{ef}
  \end{equation}
with $\beta$ and $\psi_\infty$ having the standard meaning in the
Ginzburg-Landau theory \cite{Tinkham}.

The proof of (\ref{ef}) parallels the proof of the Bernoulli potential.
The Cooper pairs move according to the Schr\"odinger equation,
  \begin{equation}  
  {1\over 2m^*}\left(i\hbar\nabla+e^*{\bf A}\right)^2\psi
    +e^*\varphi\psi=0 \,,
  \label{Se}
  \end{equation}
which replaces the force balance equation (\ref{Ne}).

The electrostatic potential $\varphi$ also causes a mismatch between
the local Fermi level and the global chemical potential that results
in a local suppression of Cooper pairs. Briefly, the charge densities
of superconducting and normal electrons have a ratio different from
the unperturbed state, $\rho_{\rm s}:\rho_{\rm n} \ne 
\rho_{\rm s}^\infty:\rho_{\rm n}^\infty$, where the superscript 
$\infty$ denotes values far from the perturbation. Due to this 
mismatch, the share of normal electrons is locally increased. In the 
linear approximation one has
  \begin{equation}  
  \rho_{\rm n}=\rho_{\rm n}^\infty-{e^2\over D}\varphi \,.
  \label{chi}
  \end{equation}
Taking the gradient, one can read (\ref{chi}) as balance between the
electric force and a kind of osmotic pressure of normal electrons, 
$e{\bf E} = D\nabla(\rho_{\rm_n}/e)$. In this sense, equation (\ref{chi})
describes the response of normal electrons to the electric field,
importance of which has already been noticed by London \cite{London}.
The ``diffusion coefficient'' $D$ is identified below.

Finally, the electrostatic potential obeys the Poisson equation,
  \begin{equation}  
    -\varepsilon\nabla^2\varphi=\rho=\rho_{\rm s}+\rho_{\rm n}+
  \rho_{\rm lattice} \,.
  \label{Pep}
  \end{equation}
With the wave function normalized to the charge density of the
condensate, $\rho_{\rm s}=e^*|\psi|^2$, we can use the condition of
neutrality,
$e^*|\psi_\infty|^2+\rho_{\rm n}^\infty+\rho_{\rm lattice}=0$,
to eliminate the normal density from the Poisson equation,
  \begin{equation}  
    -\varepsilon\nabla^2\varphi=e^*\left(|\psi|^2-|\psi_\infty|^2\right)
    -{e^2\over D}\varphi \,.
  \label{Pe}
  \end{equation}
The set of equations (\ref{Se}) and (\ref{Pe}) is closed by the
Maxwell equation, $\nabla\times\nabla\times{\bf A}=\mu{\bf j}$, with
the superconducting current ${\bf j} = -
{e^*\over m^*}{\rm Re}\,[\bar\psi(i\hbar \nabla+e^*{\bf A})\psi]$.

Let us suppose that the system is nearly neutral so that the left hand
side of (\ref{Pe}) may be disregarded. In this case, the potential
can be determined from the suppression of the superconducting density,
  \begin{equation}  
  \varphi\approx {2D\over e}\left(|\psi|^2-|\psi_\infty|^2\right) \,.
  \label{ci}
  \end{equation}
Substitution of (\ref{ci}) into the Schr\"odinder equation (\ref{Se})
yields a closed equation for $\psi$,
  \begin{equation}  
  {1\over 2m^*}\left(i\hbar\nabla+e^*{\bf A}\right)^2\psi
  + 4D|\psi|^2\psi -4D|\psi_\infty|^2\psi=0 \,.
  \label{GLeD}
\end{equation}
One can see that for $4D=\beta$ and the standard GL relation,
$\alpha=-\beta |\psi_\infty|^2$, Eq.~(\ref{GLeD}) turns into the
 GL equation,
  \begin{equation}  
  {1\over 2m^*}\left(i\hbar\nabla+e^*{\bf A}\right)^2\psi
  + \beta|\psi|^2\psi + \alpha\psi=0  \,.
  \label{GLe}
  \end{equation}
With $D=\beta/4$, Eq.~(\ref{ci}) coincides with Eq.~(\ref{ef}).

Now we show that the charge neutrality is a good approximation in
the general case. A simple estimate shows that the deviation from
charge neutrality, $\rho=-\varepsilon\nabla^2\varphi$, is much
smaller than the increase of the normal electron charge density 
$-4e^2\beta^{-1}\varphi$. Indeed, the potential varies on the scale 
of the coherence length, $\nabla^2\varphi\sim\xi^{-2}\varphi$, 
therefore the ratio of these two charge densities behaves as
  \begin{equation}  
  {\varepsilon\nabla^2\varphi\over 4e^2\beta^{-1}\varphi}\sim{\beta
  \rho_{\rm s}^\infty\kappa^2\over emc^2}\sim{k_{\rm B}^2T_c^2\kappa^2
  \over \epsilon_Fmc^2}{|\rho_{\rm s}^\infty|\over|\rho_{\rm lattice}|}
  \sim{\hbar^2\lambda^2\over m^2c^2\xi^4}.
  \label{neu}
  \end{equation}
Here we have used the BCS value of $\beta$ \cite{Gorkov}, see
Eq.~(\ref{G}), and introduced the GL parameter $\kappa=\lambda/\xi$.
Even for extreme values of the high-$T_c$ materials
[$\kappa\sim 100$, $T_c\sim 90$~K, the Fermi energy
$\epsilon_F\sim 0.08$~eV, $m\sim 4 m_e$, $\varepsilon=4 \varepsilon_0$,
$\rho_{\rm s}^\infty\sim -\rho_{\rm lattice}$] the estimate
(\ref{neu}) gives about $10^{-5}$. For conventional superconductors
the ratio (\ref{neu}) is typically about $10^{-9}$. Accordingly, the
deviation from the charge neutrality may be disregarded when one solves
for the wave function. With $\varepsilon\nabla^2\varphi\approx 0$ the
Poisson equation (\ref{Pe}) simplifies to Eq.~(\ref{ef}).

Now we compare the above result with alternative approaches. First,
far from the core one can make the local approximation of the
Ginzburg-Landau equation (\ref{GLe})
  \begin{equation}  
  {m^*v^2\over 2}\psi+\beta|\psi|^2\psi+\alpha\psi=0,
  \label{lGLe}
  \end{equation}
where the velocity is given by the current,
${\bf j}=e^*{\bf v}|\psi|^2$.
One can see that in this limit the electrostatic potential found from
(\ref{lGLe}) and (\ref{ci}) equals the Bernoulli potential (\ref{Bp}). 
We note that the incompressible superconducting liquid discussed by 
London \cite{London} corresponds to the limit $\kappa\to\infty$ of the 
GL theory \cite{Tinkham}. In this limit, the local approximation is 
exact.

Second, formula (\ref{ef}) is similar to the formula proposed by 
Khomskii and Kusmartsev \cite{KhomKusm,KhomFrei}, $\varphi_{\rm KK}=
\Delta^2/(4e\epsilon_F)$. It turns out, however, that from the 
BCS-Gor'kov relations \cite{Gorkov},
  \begin{equation}  
  \beta={24(\pi k_{\rm B}T_c)^2\over 7\zeta(3)\epsilon_FN},
  ~~~~~~~~~
  \psi={\sqrt{7\zeta(3)N}\over 4\pi k_{\rm B}T_c}\Delta \,,
  \label{G}
  \end{equation}
one obtains
  \begin{equation}  
  {\beta\over 2e}|\psi|^2={3\Delta^2\over 4e\epsilon_F} \,,
  \label{Df}
  \end{equation}
i.e., our value is three times larger than $\varphi_{\rm KK}$. 
The similarity of (\ref{Df}) with the potential of Khomskii and 
Kusmartsev is, however, accidental. While (\ref{Df}) follows from
the intertial and Lorentz forces, the potential $\varphi_{\rm KK}$
results from the condensation energy of the pairing. We have 
neglected the condensation energy in our treatment.

In the rest of this paper we show actual charge profiles of an
isolated vortex and of the vortex lattice. We use known solutions of
the Ginzburg-Landau equation and evaluate the electrostatic potential
$\varphi$ from Eq.~(\ref{ef}). The deviation of the charge density
from neutrality then follows from the Poisson equation,
$\rho = -\varepsilon\nabla^2\varphi$.

We start the discussion using the isolated vortex model suggested
by Clem \cite{Clem},
  \begin{equation}  
  |\psi|^2=|\psi_\infty|^2{r^2\over r^2+\xi_{\rm v}^2} \,,
  \label{Ca}
  \end{equation}
where $\xi_{\rm v}\sim\xi$ is given by the minimum of the free energy
as $\xi_{\rm v} = \xi\sqrt{2}
\sqrt{1-K_0^2(\xi_{\rm v}/\lambda)/K_1^2(\xi_{\rm v}/\lambda)}$,
where $K_0$ and $K_1$ are modified Bessell functions.
This wave function fails at long distances, $r \sim \lambda$,
where the correct behavior follows the local approximation,
$|\psi|^2 = |\psi_\infty|^2(1+{m^*v^2\over 2\alpha})$, but it
provides a good approximation in the vicinity of the core, $r\sim\xi$,
where the dominant charge contribution is located. The electric field,
${\bf E}=-\nabla \varphi$, obtained from (\ref{Ca}) and (\ref{ef})
is directed radially,
  \begin{equation}  
  {\bf E}=-{\beta\over e}|\psi_\infty|^2{\xi_{\rm v}^2\over(r^2+
  \xi_{\rm v}^2)^2}{\bf r} \,.
  \label{Ce}
  \end{equation}

The charge density corresponding to (\ref{Ca}),
  \begin{eqnarray}  
  \rho&=&-2\varepsilon{\beta\over e}|\psi_\infty|^2
  {\xi_{\rm v}^2(\xi_{\rm v}^2-r^2)\over(r^2+\xi_{\rm v}^2)^3}
  \nonumber\\
  &=&-{a_{\rm B}e\over\pi}{\xi_{\rm v}^2\over 2\xi^2}
  {\xi_{\rm v}^2-r^2\over(r^2+\xi_{\rm v}^2)^3},
  \label{Cc}
  \end{eqnarray}
is depleted in the core, $r < \xi_{\rm v}$, and enhanced outside the
core, $r > \xi_{\rm v}$. Here,
$a_{\rm B}={4\pi\varepsilon\hbar^2\over me^2}$ is the Bohr radius in
the crystal. 

The second form of (\ref{Cc}) can be compared with the result of
Blatter {\em et al.} \cite{Blatter}, who found a more complicated
factor, $\rho_{\rm Bl}\approx\rho\times\pi^{-2}{dT_c\over 
d\ln\epsilon_F}$. We note that Blatter {\em et al.} approached the
problem from the microscopic side using the BCS relation between the
total charge and the chemical potential. This charge is moreover
screened in the spirit of Jakeman and Pike \cite{Jakeman} within 
the Thomas-Fermi approximation. Since ${dT_c\over d\ln\epsilon_F}
\approx\ln(\hbar\omega_D/T_c)\sim$~1~to~10, their numerical results 
are close to ours.

The charge of the vortex core per unit length,
$Q=2\pi\int_0^{\xi_{\rm v}}r\rho{\rm d}r =
2\pi\varepsilon E(\xi_{\rm v}) \xi_{\rm v}$, results as
  \begin{equation}  
  Q=-\pi\varepsilon{\beta\over 2e}|\psi_\infty|^2=
  {\pi\varepsilon\alpha\over 2e}=-{a_{\rm B}e\over 4\xi^2} \,.
  \label{Cvc}
  \end{equation}
In a single pancake (a vortex of length $d=1.17\cdot 10^{-9}$~m) in 
$\rm YBa_2Cu_3O_{7-\delta}$ (YBCO)
at 77~K, the depleted charge is $Q*d=-0.004$~e. We have used
$a_{\rm B}=0.05$~nm and $\xi=1.91$~nm, where the coherence lenght follows
from $\xi=\lambda/\kappa$ with $\lambda=191$~nm and $\kappa=100$. With
the quasiparticle screening in the spirit of van~Vijfeijken and 
Staas\cite{van}, this value will be reduced by 
$\rho_{\rm s}/\rho=2|\psi_\infty|^2/N\approx 1-(T/T_c)^4=0.5$ for 
$T=77$~K.

Note that the vortex charge (\ref{Cvc}) has a value independent of the core
diameter $\xi_{\rm v}$. A similar value, $Q=-1.28\times a_{\rm B}e/(4\xi^2)$,
results for a different model function, $|\psi|=|\psi_\infty|\tanh(r/\xi)$,
recommended in \cite{Tinkham}. This seems to contradict the expectation from
the local approximation, where the core diameter plays a dominant role. In
the local approach, the velocity close to the core reads
$v\approx{e\Phi_0\over 2\pi mr}$. From the Bernoulli potential (\ref{Bp})
and the Poisson equation one finds that inside the cylinder of radius $r$
the local approximation predicts a charge
  \begin{equation} 
  Q_{\rm la}=-{e\varepsilon\Phi_0^2\over 2\pi m r^2}.
  \label{la}
  \end{equation}
With the core radius equal to the coherence length, 
\mbox{$r=\xi=\hbar/\sqrt{-2m^*\alpha}$}, the classical estimate (\ref{la})
reproduces the nonlocal result (\ref{Cvc}). Although the exact agreement
of Eq.~(\ref{la}) with Eq.~(\ref{Cvc}) is rather accidental, it shows
that classical estimates used in the literature are reasonable.

\begin{figure}[h]  
  \centerline{\parbox[t]{8cm}{
  \psfig{figure=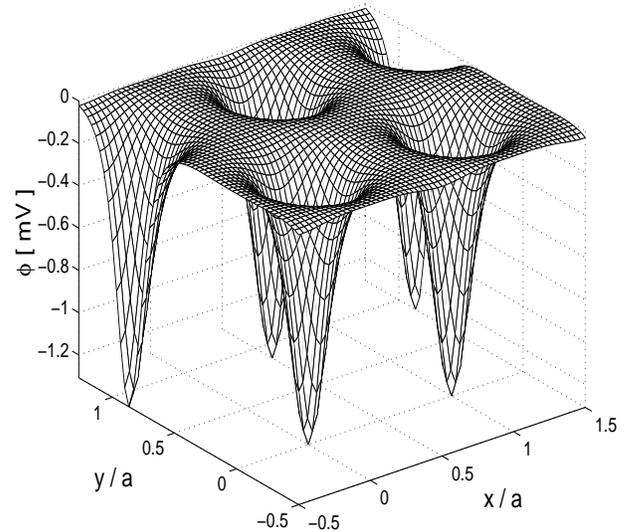,width=8cm,height=7cm}}}
\vskip 2pt
\caption{Electrostatic potential $\varphi$ in a superconductor with
triangular vortex lattice of spacing $a$. The shape of the potential
follows the magnitude of the GL function, see Eq.~(\protect\ref{ef}),
which depends
exclusively on the reduced induction $b=B/B_{c2}=0.05$ and the GL
parameter $\kappa=100$. The presented magnitude of $\varphi$ results
for $\lambda=191$~nm and $m=4m_0$, which corresponds to YBCO at
temperature  77~K.}
\label{f1}
\end{figure}

An example of the electrostatic potential for the triangular vortex
lattice is presented in Fig.~\ref{f1}. One can see that in the core
region the potential has a shape similar to the Clem model. The
maximum amplitude, $\varphi_{\rm max} = -1.3$~mV, is identical to the
maximum potential obtained from the Clem model since the GL function
$|\psi|^2$ reaches zero at the vortex center in both cases.
The GL function used in Fig.~\ref{f1} has been evaluated by
the iteration method described in \cite{ehb2}.

  \begin{figure}[h]   
  \centerline{\parbox[c]{8cm}{
  \psfig{figure=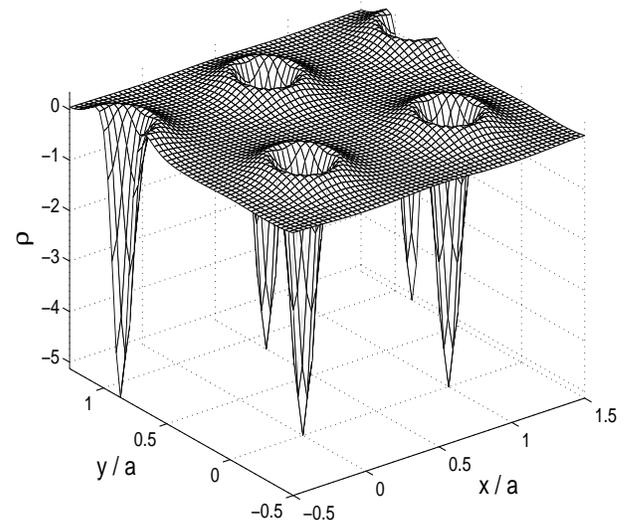,width=8cm,height=7cm}}}
\caption{Charge density $\rho = -\varepsilon \nabla^2 \varphi$
in a superconductor with a triangular vortex
lattice. All the parameters are the same as in Fig.~\protect\ref{f1}.
The charge density is plotted in units $10^{-6}$ electrons per unit
cell, with the volume of the unit cell $0.173$~nm$^3$ corresponding
to YBCO.}
\label{f2}
\end{figure}

The density of charge corresponding to the potential in Fig.~\ref{f1}
is shown in Fig.~\ref{f2}. In the core, the charge is negative, i.e.,
the density of charge carriers is depleted. Outside the core, the charge
is positive, i.e. there is charge built up which screens the charge of
the core. The radius of the vortex core (radius at which the charge
changes sign) found in Fig.~\ref{f2} is $1.53\xi$, comparable to the
value $1.41\xi$  obtained from the Clem model. The maximum of the
screening charge appears at $2.4\xi$, while the Clem model yields
$2\xi$. The maximum of the  depleted charge in the core centre is
$-5 \cdot 10^{-6}$~electron/cell while  the Clem model gives
$-6.8 \cdot 10^{-6}$~electron/cell. The core charge per unit lenght
is thus reduced to 85\% of the Clem model. These numerical
results thus show that the approximation based on the Clem model is
reasonable inside and in the vicinity of the core.

In conclusion, we have shown that the GL function can be used to
evaluate the electric field in superconductors, see Eq.~(\ref{ef}), in
spite of the fact that the electrostatic potential is absent in the
GL theory. For convenient estimates of the core charge we have
derived the simple formula (\ref{Cvc}). Our theory neglects 
contributions from the condensation energy and the screening by normal 
particles; more work on this is under way. Due to the neglected 
screening, the theory applies only for intermediate temperatures 
where the GL theory still gives reasonable results while the share 
of the superconducting electrons is already of the order of unity. 
Numerical results relevant to YBCO at the temperature 77~K and at low 
magnetic field with well separated vortex cores have been presented.

\medskip
This work was supported by M\v{S}MT program 
Kontakt ME160 and GA\v{C}R 202000643,
GAAV A1010806 and A1010919 grants. 
The European ESF program VORTEX is gratefully acknowledged.

\end{document}